\documentclass[preprint,nofootinbib]{revtex4}
\usepackage[dvips]{color}
\usepackage[dvips]{graphicx}
\usepackage{epsfig}
\usepackage{subfig}

\begin{document}

\title{Testing cosmic isotropy with galaxies position angles distribution}

\author{R. S. Menezes Jr.$^{1}$\footnote{rsmjr@ifba.edu.br}, C. Pigozzo$^2$\footnote{cpigozzo@ufba.br} and S. Carneiro$^{2}$\footnote{saulo.carneiro@pq.cnpq.br}}

\affiliation{$^{1}$Departamento de F\'{\i}sica, Instituto Federal da Bahia, 40301-015 Salvador, BA, Brazil\\ $^{2}$Instituto de  F\'{\i}sica, Universidade Federal da Bahia, 40210-340 Salvador, BA, Brazil}

\date{\today}

\begin{abstract}
We analyse the distribution of position angles of 1 million galaxies from the Hyperleda catalogue, a sample that presents the galaxies coordinates in the celestial sphere, information that allows us to look for a possible privileged direction. Our analysis involves different tests and statistical methods, from which it is possible to infer with high probability ($p$-value extremely low) that the galactic planes are not randomly oriented in the sky. Whether this is an evidence of a cosmological anisotropy or an observational bias due to local effects is something deserving further studies.
\end{abstract}

\maketitle

\section{Introduction}

The Cosmological Principle is one of the basis of Cosmology, relying on precise observations like the highly isotropic cosmic background radiation and the high-level isotropy observed in the distribution of cosmic structures at scales above two hundred  meparsecs. This fundamental principle is an inherent part of the standard model of our cosmos, through the Robertson-Walker metric on which it is builded. Like any other assumption behind our modern description of the Universe, the Cosmological Principle has been always under test, in order to be precisely verified in different scales of distance and redshift. 

Since the 1980 decade possible signatures of a cosmic anisotropy began to appear. In 1982, P. Birch has found an apparent anisotropy in the relation between the polarization vectors of the magnetic field of the radiation received from galaxies and their respective position angles (PA) \cite{Birch}. In that paper, he has attributed the discovered anisotropy to a cosmic rotation and estimated the angular velocity and the direction of the anisotropy axis as  $10^{-13}$ \ rad/year and $(\alpha_p; \delta_p) = (44^\circ; 35^\circ)$, respectively. This hypothesis of a rotating Universe had already been thought by Gamow in 1946 and 1952 \cite{gamow1946,gamow1952} as a possible explanation for the galaxies rotation (see also \cite{li,Carneiro}). In later years, D. Kendall and G. Young \cite{kendall_young} and Andreasyan \cite{andreasyan} obtained results similar to Birch's.
In 1997, J. Ralston and B. Nodland, observing the polarised radiation from galaxies, have encountered again a preferential direction ($(\alpha_p; \delta_p) = (135^\circ; 0^\circ)$) and they also have attributed it to a cosmic rotation, excluding the possibility of this anisotropy to be due to the Faraday effect \cite{ralston_nodland}.
Some further evidences of anisotropy have also appeared in observations of
supernovae \cite{antoniou_perivolaropoulos, kalus, yang_etal, gupta, cai_tuo},
bulk flow \cite{watkins_etal},
fine structure constant in quasars \cite{webb_etal, mariano_perivolaropoulos},
quasars alignment \cite{hutsemekers1998, hutsemekers2001, hutsemekers2005, hutsemekers2014, cabanac},
and the CMB dipole, quadrupole and octopole \cite{lineweaver, tegmark_etal, land_magueijo, frommert_ensslin, bielewicz_etal}. In 2011, Koivisto {\it et al.} showed that an anisotropic model was compatible with observations of supernovae \cite{koivisto}. We have obtained, in 2013, similar results with another sample of supernovae and also using the observed position of the first acoustic peak of the CMB anisotropy spectrum \cite{menezes_etal}. Probes of the cosmic isotropy with type Ia supernovae can also be found, for example, in \cite{welber,jailson}.

All these potential evidences motivate the search of new observational anisotropy signatures. Our goal in this paper is to look for anisotropies in galaxies distributions, more precisely in the distributions of their position angles (PA). If there is a cosmic anisotropy related to a cosmic rotation, the PA distribution will not be uniform, once the rotation will affect the PA orientations of the galaxies, with an alignment of the galactic planes generated by the ``centrifugal force" \cite{li}. For this purpose, we will use the Hyperleda catalogue, a sample containing approximately 1 million galaxies  \cite{hyperleda}\footnote{http://vizier.cfa.harvard.edu/index.gml}. This catalogue, besides of containing information on position angles, presents the coordinates of each galaxy in the celestial sphere, information that can allow us to identify a possible privileged direction. Our analysis will involve different tests, that will be presented in the next sections along with the corresponding results.

\section{Uniformity test}

The first test to be accomplished is the uniformity test. It shows how close a dataset approaches a random distribution. The random distribution in this case will correspond to a uniform distribution, that is, with no bias in the dataset. This will be our null hypothesis.
In order to do the test, we define the $\chi^2$ function
\begin{equation}\label{chi2_uniform1}
  \chi^2 \equiv \sum_{i=1}^{C} \frac{(f_i-\bar{f})^2}{\bar{f}^2},
\end{equation}
where $C$ is the number of categories, $f_i$ is the frequency in which a given event occurs and $\bar{f}$ is the mean frequency. In our case, we divided the PA in $C=180$ categories corresponding to intervals of $1^\circ$ ($[0^\circ,1^\circ),[1^\circ, 2^\circ),..., [179^\circ, 180^\circ)$), because the circular data are bimodal (the $0^\circ$ slope is equal to the $180^\circ$ slope). It is easy to note that a perfect uniform distribution corresponds to $\chi^2=0$, thus, the smaller the value of $\chi^2$, the uniformer the distribution.
Hyperleda catalogue contains  $833,844$ galaxies and, with their PA data, the frequencies vary between $f = 4,170$ and $f =5,039$. Applying equation (\ref{chi2_uniform1}) to this dataset we find
\begin{equation}\label{chi2_uniform1_result}
  \chi^2_* = 919.042. \nonumber
\end{equation}

In order to verify how high or low $\chi^2$ should be to permit deciding whether the null hypothesis is true or false, we can establish a reference probability $p$, calculated from a Gaussian distribution, for the number of degrees of freedom $M = C - 1$, which we get as a reference value. The criteria commonly adopted for rejecting the null hypothesis is that $p < 5\%$. This criteria establishes that the probability of a distribution to be due to chance is less than $5\%$ (the bias hypothesis, therefore, is greater than $95\%$). Small values of $\chi^2$ correspond to the uniformity hypothesis, then it is expected that more than $95\%$ of the data belongs to the $\chi^2 > 919.042$ interval for a uniform distribution.

\begin{figure}[t]
  \centering
  \includegraphics[height=5.5cm]{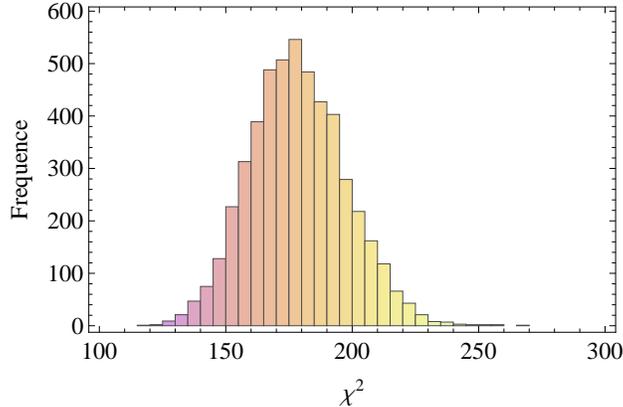}\\
  \caption{\small{Histogram of $\chi^2$ obtained with 5,000 random samples of position angles.}}\label{result_uniform_test}
\end{figure}

We have randomly generated $5,000$ samples containing $833,844$ position angles and, for each generated distribution, we calculated the corresponding $\chi^2$. The result is shown in Fig. \ref{result_uniform_test}. We can see that the value obtained for the real catalogue, $\chi^2_* = 919.042$, is much larger than the upper limit of the $\chi^2$ distribution obtained with the PA random distributions. Therefore, the actual PA distribution is not that expected for an aleatory distribution.

\section{Medium alignment - $r$ test}

This test consists in determining the value of a parameter $r$ ($0 \leq r \leq 1$) that quantifies how near the uniformity a directional data sample is. We wish to compare it with the values found for $5,000$ samples randomly generated. As we will see, $r \rightarrow 1$ for samples with a high degree of alignment, while $r \rightarrow 0$ when the alignment is null.

Let us consider a sample of $N$ directional data. Each direction in the sample can be represented by a unitary vector $\vec{v}_i$ that form an angle $\theta_i$ with a reference direction. We can, therefore, define the vector $\vec{R} = \Sigma \vec{v}_i$, with components
\begin{eqnarray}
  \label{C1} C_1 &\equiv& \sum_{i=1}^{N}\cos\theta_i, \\
  \label{S1} S_1 &\equiv& \sum_{i=1}^{N}\sin\theta_i.
\end{eqnarray}
However, owing to the bimodal nature of PA data, the angles must be multiplied by $2$ \cite{mardia_jupp}, and hence (\ref{C1}) and (\ref{S1}) are corrected to
\begin{eqnarray}
  \label{C2} C_2 &\equiv& \sum_{i=1}^{N}\cos2\theta_i, \\
  \label{S2} S_2 &\equiv& \sum_{i=1}^{N}\sin2\theta_i,
\end{eqnarray}
and we have
\begin{equation}\label{R}
R = \sqrt{{C_2}^2+{S_2}^2}.
\end{equation}
We now define
\begin{equation}\label{r}
\frac{R}{N} = \frac{\sqrt{{C_2}^2+{S_2}^2}}{N} = \sqrt{{c_2}^2+{s_2}^2} \equiv r,
\end{equation}
where
\begin{eqnarray}
  c_2 & \equiv & \frac{C_2}{N} = \frac{\sum_{i=1}^{N}\cos2\theta_i}{N}, \\
  s_2 & \equiv & \frac{S_2}{N} = \frac{\sum_{i=1}^{N}\sin2\theta_i}{N}.
\end{eqnarray}
The direction $\bar{\theta}$ is defined by the relations
\begin{eqnarray}
  \label{cos_mean} \cos\bar{\theta} &=& \frac{c_2}{r}, \\
  \label{sin_mean} \sin\bar{\theta} &=& \frac{s_2}{r},
\end{eqnarray}
or, alternatively, by the equation
\begin{equation}\label{theta_median}
  \bar{\theta} = \arctan\left(\frac{s_2}{c_2}\right),
\end{equation}
noting that the signs of (\ref{cos_mean}) and (\ref{sin_mean}) are important to determine the quadrant of $\bar{\theta}$.

\begin{figure}[t]
\centering
{\includegraphics[height=4.5cm]{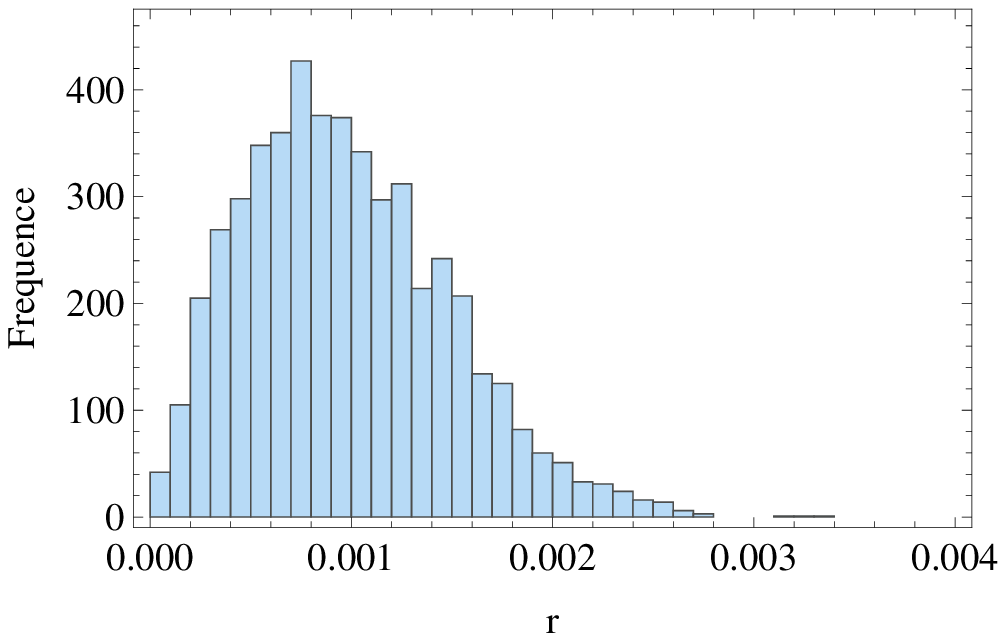}}
{\includegraphics[height=4.5cm]{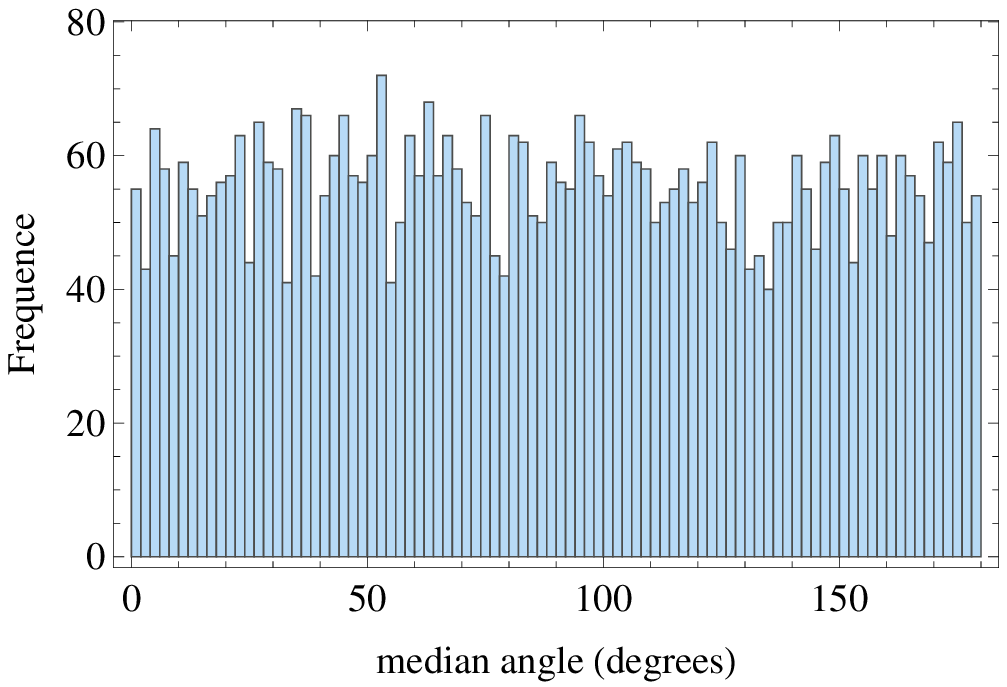}}
\caption{\small{{\bf Left panel:} Histogram of $r$  for $5,000$ PA random samples. \newline {\bf Right panel:} The corresponding median angles distribution.}}\label{histr}
\end{figure}

Once we have defined our parameters, the $r$-test consists in calculating the value $r^*$ for the position angles $\theta$ of the $833,844$ galaxies of Hyperleda and comparing them with the values found for $5,000$ samples of galaxies randomly generated. In this way, we have obtained
\begin{equation}\label{r_result}
  r^* = 0.0044 \nonumber,
\end{equation}
and for the median angle we have
\begin{equation}\label{theta}
  \bar{\theta} = 177^\circ \nonumber.
\end{equation}
This value of $r^*$ seems too low for inferring the existence of any alignment. However, after generating random samples, we have obtained the histogram shown in the left panel of Fig. \ref{histr}. We can see that, for all generated samples, $r < r^*$. This means that the probability of obtaining a value  $r \geq r^*$ from a random distribution is low. In the right panel of Fig. \ref{histr} we see that the median angles obtained from the random samples obey an approximately uniform distribution.

Despite of the fact that $r < r^*$ for all randomly generated samples, we should be causious before saying that this indicates the existence of a global alignment of the galactic planes. In fact, the existence of a local alignment would also caracterise an anisotropy. In order to verify this possibility, we will perform the following test.

\section{Local medium alignment - $r$ value}

We now wish to test the directional alignment. For this purpose, let us calculate the value of $r$ for a given position in the sky. We divide the celestial sphere in angular cells with $\Delta l = 30^\circ$ and $\Delta b = 15^\circ$ (in galactic coordinates) and calculate $r$ for each of them. These values were chosen so that each cell would contain a significative number of galaxies - in our case, the number varies between $125$ and $19,508$ galaxies. A thinner  partition would result in too few galaxies in some cells, artificially producing a strong alignment (for example, the choice $\Delta l = 15^\circ$ e $\Delta b = 7.5^\circ$ has lead to cells with just one or two galaxies).

\begin{figure}[t]
  \centering
  \includegraphics[width=9cm]{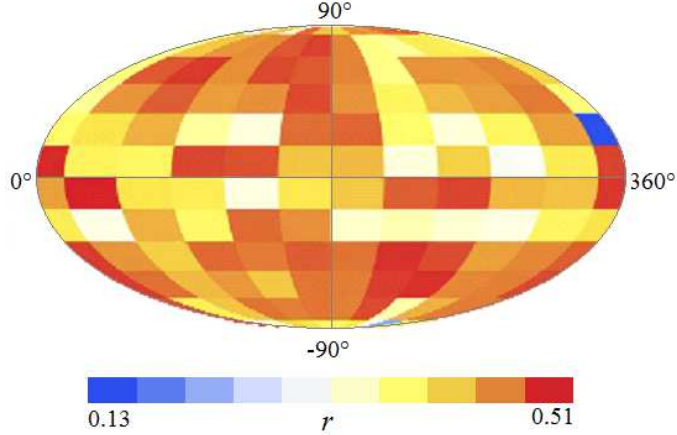}\\
  \caption{Values of $r$ for different directions.}\label{dist_angular_r}
\end{figure}

Fig.  \ref{dist_angular_r} shows the obtained result. The region with stronger alignment is that with ($l,b$) $\in$ ($[0^\circ,30^\circ], [0^\circ , 15^\circ]$), in which $r=0.507$. Around 87.5\% of the values found for $r$ are above $0.385$, the central value of the distribution. Most of the cells presents values $0.4 < r < 0.5$, which characterises a homogeneous distribution. However, we can observe two approximately bimodal regions with higher $r$ (redder in the figure). These regions correspond to the intervals ($l,b$) $\in$ ($[210^\circ,270^\circ],[-30^\circ,-75^\circ]$) for the galactic southern hemisphere, and ($l,b$) $\in$  ($[120^\circ,180^\circ],[30^\circ,75^\circ]$) for the northern. The number of galaxies in these two regions varies between $8,094$ and $14,391$ for the southern hemisphere, and between $5,700$ and $6,996$ for the northern. In both hemispheres we have $0.49 < r < 0.50$, higher values when compared to the other regions.

In this test our goal was to verify how strong are the alignments between galaxies in different regions of the sky. In next section we will find the mean direction of the position angles of galaxies belonging to these same regions and compare to their neighborhood.

\section{Local medium alignment - $\bar\theta$ value}

\begin{figure}
  \centering
  \includegraphics[width=9cm]{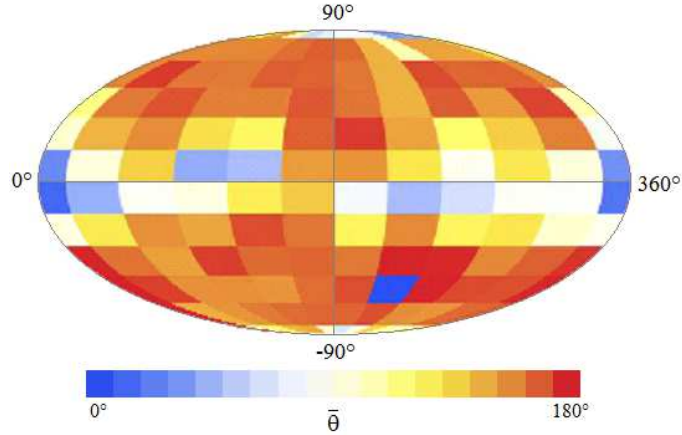}\\
  \caption{Values of the median angle $\bar{\theta}$ for different directions.}\label{dist_angular_teta}
\end{figure}

The median angle of the directional distribution can be obtained from equation (\ref{theta_median}). Let us derive its value for the same regions previously defined and compare the results with those from last section. The result is shown in Fig. \ref{dist_angular_teta}. It is possible to see that the position angles spam all the interval from $0^\circ$ to $180^\circ$, but with a predominance of $\bar{\theta}>150^\circ$ (reddest regions), specially regions outside the Milk Way galactic plane ($b=0^\circ$). In the regions of predominance, on the other hand, the distribution presents some uniformity. The region ($l,b$) $\in$ ($[210^\circ,270^\circ],[-30^\circ,-60^\circ]$) is worthy of note, where angles  $\sim 180^\circ$ are strongly concentrated. This region coincides with one of the cells with highest values of $r$ found previously\footnote{As position angles are axial data, the inclination $\bar{\theta} \sim 180^\circ$ (in red) coincides with $\bar{\theta} \sim 0^\circ$ (blue).}. In this same direction, but in the opposite side, we can observe a significative uniformity in the values of $\bar{\theta}$, which rest in the interval  $\bar{\theta} \in [160^\circ,180^\circ]$ for the region  ($l,b$) $\in$ ($[0^\circ,180^\circ],[30^\circ,75^\circ])$. This indicates a remarkable concordance between the median angles on these regions. Nevertheless, this concordance cannot, alone, be taken as a signature of alignment, since low values of $r$ mean weak alignment inside a given region. Our results for the median position angles should be interpreted in the light of the previous $r$-test. We can conclude that the aligned cells identified in the $r$-test coincide with the regions
\begin{equation}\label{eixo_alinhamento}
  (l,b) \in ([210^\circ,270^\circ],[-30^\circ,-60^\circ]) \quad \textrm{and} \quad  (l,b) \in ([120^\circ,180^\circ],[45^\circ,75^\circ]).\nonumber
\end{equation}

\section{Modified $S_D$ test}

The $S_D$ test is an statistical test designed for sets of directional data. Consider a set composed by $N$ axial data represented by angles defined with relation to some reference frame. The difference between two angles of this distribution, $\alpha$ and $\beta$, can be determined by the relations
\begin{equation}\label{sinx}
  (\alpha - \beta)=  \left\{\begin{array}{ll}
        \frac{\pi}{2} - \left(\frac{\pi}{2} - |\alpha - \beta|\right), \quad &\textrm{if} \quad  |\alpha - \beta| \leq \pi/2 ,\\
        \frac{\pi}{2} - \left|\frac{\pi}{2} - |\alpha - \beta|\right|, \quad  &\textrm{if} \quad  |\alpha - \beta| > \pi/2,\\
    \end{array}
\right.
\end{equation}
which can be rewritten in the general form
\begin{equation}
  (\alpha - \beta) = 90^\circ - |90^\circ - |\alpha - \beta||.
\end{equation}
Therefore, the dispersion of a set of directional angular data $\theta_1, \theta_2, ..., \theta_N$ around a given angle (also known as circular average standard deviation) is given by \cite{mardia_jupp}
\begin{equation}\label{dispersão}
   D(\tilde{\theta}) = \frac{1}{N}\sum_{i=1}^{N}\{90^\circ - |90^\circ - |\theta_i - \tilde{\theta}||\} = 90^\circ - \frac{1}{N}\sum_{i=1}^{N}|90^\circ - |\theta_i - \tilde{\theta}||,
\end{equation}
where $\tilde{\theta}$ is the angle that minimises the dispersion $D$, called the sample median angle.

Let us now consider $n_v$ galaxies in the vicinity of a $j$-th galaxy. The dispersion of $\theta$ for this set is 
\begin{equation}\label{dispersão1}
   D_j(\tilde{\theta}) =  90^\circ - \frac{1}{n_v}\sum_{i=1}^{n_v}|90^\circ - |\theta_i - \tilde{\theta_j}||.
\end{equation}
If we calculate this dispersion for all the $N$ galaxies of our set, we can define the mean dispersion
\begin{equation}\label{SD}
  S_D = \frac{1}{N}\sum_{j=1}^{N}D_j.
\end{equation}
Therefore, $S_D$ measures the level of angles concentration for objects relatively near in space. It tends to be smaller in distributions of directional data with some alignment, as compared to those where the angles are randomly distributed. Hence, the value of $S_D$ can be used as an indicator of the presence of an alignment in the distribution.

\begin{figure}
  \centering
  \includegraphics[width=9cm]{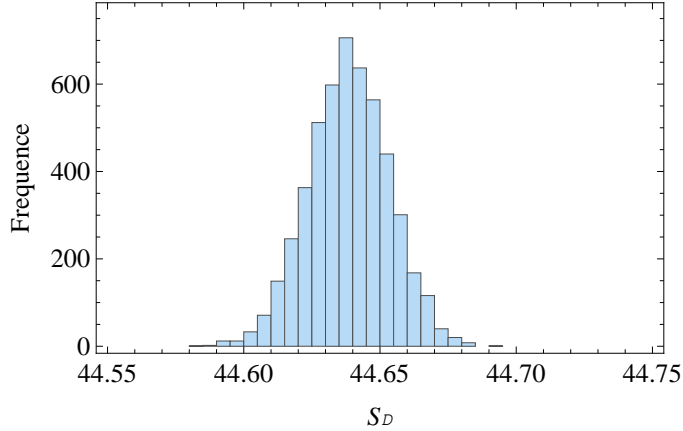}\\
  \caption{Histogram of $S_D$ for $5,000$ PA random samples.}\label{hist_SD_galaxias}
\end{figure}

The test presented above involves, however, some computational issues, because the used sample contains around $10^6$ galaxies. Selecting the number of neighbors of a given galaxy involves to calculate the distance from such a galaxy to all the other galaxies in the sample, and then repeat this procedure for each galaxy in the set. Another difficulty resides in the definition of $n_v$, the number of neighbors. For these reasons, we proceed to previously select a region in the sky and count the $n_g$ galaxies inside it. This will be the number of neighbors. After that, we calculate $S_D$ (let us call it $S^*_D$) and compare it with the values obtained from $5,000$ random samples of position angles. The cells in the sky were selected as in the previous tests, that is, $\Delta l = 30^\circ$ and $\Delta b = 15^\circ$. In this way, in the modified $S_D$ test we:
\begin{enumerate}
  \item select a cell in the sky delimited by $\Delta l= 30^\circ$ and $\Delta b = 15^\circ$;
  \item count the number of galaxies in that region, $n_g$;
  \item calculate the median  $\tilde{\theta}_j$ corresponding to these $n_g$ galaxies, by determining the minimum value of $D$ ($D_j$) when $\tilde{\theta}$ varies;
  \item repeat this procedure for all cells ($N = 144$);
  \item calculate the mean $S_D$ of all the $D_j$ found;
  \item generate a random sample of position angles;
  \item perform procedures 1 to 5 for the random sample;
  \item repeat procedures 6 and 7 for $5,000$ random samples.
\end{enumerate}
Following the steps above we have obtained
\begin{equation}
\nonumber   S^*_D= 44.19^\circ,
\end{equation}
along with the histogram shown in Fig. \ref{hist_SD_galaxias} for the simulated values. We can see that all the simulated values are above $S^*_D$, with a maximum at $S_D = 44.58^\circ$. This suggests a global alignment, in agreement to the results of the uniformity and $r$ tests of the previous sections.

\section{Conclusions}

We have studied in this paper the distribution of position angles of around 1 million galaxies belonging to the Hyperleda catalogue. Applying diverse statistical methods to the data, it was possible to infer, with a high probability ($p$-value extremely low), that these galaxies do not seem to be randomly positioned in the sky. In this way, our results point out to the possible existence of an anisotropy in the Universe. This anisotropy has appeared in the form of a non-homogeneous PA distribution in the celestial sphere, as well as in the presence of local alignments. Comparing the different statistics used, we have found stronger alignments (larger values of $r$) in the directions  ($l$, $b$)  $\in ([120^\circ,180^\circ],  [30^\circ,75^\circ]$) and ($l$, $b$) $\in ([210^\circ,270^\circ], [-30^\circ,-75^\circ]$).

Nevertheless, two issues force us to be careful in our conclusions. The first is the size of the used catalogue. One million galaxies is a significative figure, but very small when compared to estimates of the total galaxies inside the horizon, around $170$ billions \cite{gott}. Our sample represents only $0.0006\%$ of the total galaxies in the observable universe. The second issue is related to the fact that we have not considered the galaxies redshifts in our analysis. This may produce an observational bias if the anisotropy is owing to some local effect. On the other hand, possible alignments in different redshifts may be hidden by the effect of superposition of data in the celestial sphere. For all these reasons, the potential evidences presented in the previous sections in favor of the existence of a global anisotropy need to be corroborated by supplementary studies.

\section*{Acknowledgements}

We are thankful to A. Bernui, J. G. V. Miranda, M. Quartin and A. L. B. Ribeiro for useful discussions. S. Carneiro is partially supported by CNPq - the Brazilian National Council for Scientific Research, with grant \#309792/2014-2.

{}

\end{document}